\documentstyle[12pt]{article}
\begin{document}
\setlength{\baselineskip}{0.30in}
\newcommand{\beq}{\begin{equation}}
\newcommand{\eeq}{\end{equation}}
\newcommand{\bi}{\bibitem}

{\hbox to\hsize{June, 1996  \hfill TAC-1996-012}

\begin{center}
\vglue .06in
{\Large \bf {Nonequilibrium corrections to energy spectra of
massive particles in expanding universe.}
}
\bigskip
\\{\bf A.D. Dolgov}
 \\[.05in]
{\it{Teoretisk Astrofysik Center\\
 Juliane Maries Vej 30, DK-2100, Copenhagen, Denmark
\footnote{Also: ITEP, Bol. Cheremushkinskaya 25, Moscow 113259, Russia.}
}}\\[.40in]

\end{center}
\begin{abstract}
Deviations from kinetic equilibrium of massive particles caused by the universe
expansion are calculated analytically in the Boltzmann approximation. For the
case of an energy independent amplitude of elastic scattering, an exact partial
differential equation is derived instead of the usual integro-differential one.
A simple perturbative solution of the former is found. For the case of an
energy-dependent amplitude the problem cannot be reduced to the differential
equation but the solution of the original integro-differential equation can be
found in terms of the Taylor expansion, which in the case of aconstant
amplitude
shows a perfect agreement with the perturbative solution of the differential
equation. Corrections to the spectrum of (possibly) massive tau-neutrinos are
calculated. The method may be of more general interest and can be applied to
the  calculation of spectrum distortion in other (not necessarily cosmological)
nonequilibrium processes.
\end{abstract}
%\newpage
\section{Introduction}

Abundance of stable relics from the hot epoch of the universe evolution are
usually calculated with the help of the integrated in momentum Boltzmann
kinetic equation,
so that the complicated integro-differential equation is reduced to an
ordinary differential equation for the number density of particles in question.
To make this reduction one has to assume that these particles are in
kinetic equilibrium with the cosmic plasma so that their distribution is
given by the expression
\beq{
f(p,t)= \exp[(\mu(t) - E) /T(t)]
\label{fpt}
}\eeq
where $E$ is the particle energy,
$\mu(t)$ is an effective chemical potential (in total equilibrium usually
$\mu(t) \equiv 0$) and $T(t)$ is the plasma temperature. Under this assumption
the particle number density
\beq{
n(t) = g_s\int {d^3p \over (2\pi)^3}f(p,t)
\label{nt}
}\eeq
was shown to satisfy the equation \cite{zop,zn}
\beq{
\dot n + 3Hn = \langle \sigma v \rangle (n^2_{eq} -n^2)
\label{ndot}
}\eeq
where $H=\dot a /a$ is the Hubble parameter, $\langle \sigma v \rangle$ is the
thermally averaged product of the annihilation cross-section and the particle
velocity, and $n^2_{eq}$ is the equilibrium
number density obtained from
the expression (\ref{nt}) with $\mu=0$.
This equation was used in particular for the derivation of the cosmological
bound on the mass of a possible heavy neutrino \cite{vdz,lw} and for the
calculation of the mass density of heavy particles, e.g. stable
supersymmetric relics, which may constitute invisible matter
in the universe. A detailed derivation of eq. (\ref{ndot})  valid for any
form of the cross-section is given in ref.\cite{gg}
(see also the book \cite{kt}).

Usually kinetic equilibrium is maintained long after breaking of chemical
equilibrium, characterized by a nonzero $\mu$. Though elastic
and annihilation cross-sections are typically of comparable magnitude, the
rate of annihilation, $(\dot n/n)_{ann} \sim \sigma v n$,
exponentially goes down with decreasing temperature together with the
number density of
the nonrelativistic annihilating particles, $n\sim \exp (-m/T)$, while the
probability of elastic scattering, $(\dot n/n)_{el} \sim \sigma v n_0$, dies
down much slower, because $n_0 \sim T^3$. Here $n_0$ is the number density
of massless (or light particles) which elastically scatter on the heavy ones
and preserve kinetic equilibrium.

However for the particular case of tau-neutrino with the mass in MeV range,
deviations from the kinetic equilibrium may be essential and can change the
results of the calculations \cite{ktcs,dr} based on equation (\ref{ndot}).
Correspondingly the nucleosynthesis bounds on a possible mass of nu-tau would
be different. Due to a nonzero mass, $\nu_\tau$ would cool faster in the
course of expansion, than massless particles. This in turn would result in less
efficient annihilation of $\nu_\tau \bar \nu_\tau$ and
in their larger frozen number density.  Thus the nucleosynthesis
bound on the mass of $\nu_\tau$ would be stronger.
In ref.\cite{dr} a simple attempt to take into account nonequilibrium
corrections to the $\nu_\tau$-distribution in energy
was made. Namely it was assumed that
the equilibrium form of nu-tau spectrum is maintained down to a certain
temperature $T_{el}$, determined by the strength of the elastic scattering, and
below $T_{el}$ the distribution function is given by the expression valid
for noninteracting particles, $f\sim \exp[\sqrt{(m^2/T_{el}^2 + p^2/T^2)]}$.
The
effect was found to be relatively weak.

An important influence on the
primordial nucleosynthesis may produce nonequilibrium electronic neutrinos
originating from the interaction with nonequilibrium
tau-neutrinos. There are two
possible effects. First, there should be an excessive cooling of massless
neutrinos due to elastic scattering on colder-than-equilibrium
$\nu_\tau$'s. This
makes the production of $\nu_\tau$ in the collision of the massless neutrinos
(inverse annihilation) less efficient and correspondingly leads to a smaller
number density  of $\nu_\tau$. It weakens the nucleosynthesis bound. Another
effect is the distortion of $\nu_e$ spectrum due to energetic
electronic neutrinos coming
from the process $\nu_\tau \bar \nu_\tau \rightarrow \nu_e \bar \nu_e$. Such
nonequilibrium $\nu_e$'s give rise to a larger neutron-to-proton ratio.
This effect was found to be
stronger than the first one and an overall stronger bound on $m_{\nu_\tau}$
was obtained\cite{dpv}.

In ref.\cite{fko} it was assumed that massless neutrinos maintain their
canonical distribution in momentum but, as the massive ones, acquire a
nonzero chemical potentials (see eq.(\ref{fpt})). This assumption permits to
reduce the problem to the solution of a system of ordinary differential
equations. It considerably simplifies the calculations. Such overall heating
would result in a weaker nucleosynthesis bound
but the neglected spectrum
distortion have a stronger impact on the nucleosynthesis\cite{dpv}.

It was claimed in ref.\cite{hm},
on the basis of numerical solution of the complete system of the exact
integro-differential Boltzmann equations,
that deviations from kinetic equilibrium generated by massive $\nu_\tau$ are
very strong and this gives rise to a much weaker bound on the nu-tau mass
than it was previously stated\cite{ktcs,dr}. In view of the arguments
presented above the sign of the effect should be just opposite. Recently the
authors found an error in their computations (Madsen, private communication)
and the result seems to be close to those existing in the literature.

Because of the complexity of numerical calculations it may be interesting to
find an approximate analytical way to calculate spectral distortion for both
massive and massless neutrinos. In this paper the spectrum of massive
particles,
which have elastic interaction with the equilibrium bath of
massless ones, is found.
In the following section the partial differential equation
for the spectral density is derived from
integro-differential Boltzmann equation for the case of the energy independent
amplitude of elastic scattering. This equation permits a simple perturbative
solution. For the case of the energy dependent amplitude such simplification
was not found but one can solve the integro-differential equation expanding
the distortion into the Taylor series. For the case of a constant amplitude of
elastic scattering, comparison of this method with the
perturbative solution of the differential equation, derived in Sec.2, is made
in
Sec.3. The agreement is found to be very good. More practically interesting
case
of weak interactions with the amplitude proportional to particle energies
is considered in Sec.4.  The results are applied to the calculations of the
corrections to the spectrum of massive tau-neutrinos.

\section{Differential kinetic equation for a constant amplitude of elastic
scattering.}

Let us consider elastic scattering of massive particles with mass $m$,
momentum {\bf $p$}, and energy $E$ on massless ones with momentum {\bf $k$}
and energy $\omega$.
The basic Boltzmann equation in the Robertson-Walker expanding universe and
in the limit of Boltzmann statistics has the following form:
\begin{eqnarray}
 (\partial_t + Hp\partial_p)f_m =
{1\over 2E} \int{d^3k\over 2\omega (2\pi)^3}{d^3p'\over E' (2\pi)^3}
{d^3k'\over 2\omega' (2\pi)^3}|A_0|^2(2\pi)^4\delta^4(p+k-p'-k')
\nonumber \\
\left[f_m(p)f_0(k) - f_m(p')f_0(k')\right]
\label{dtfm}
\end{eqnarray}
Massless particles are assumed to be in thermal equilibrium so that
$f_0(\omega)=\exp(-\omega/T)$. In the case considered in this section, the
amplitude of elastic scattering $A_0$ does not depend on the momenta of the
colliding particles. All the integration but one in this expression can be
explicitly done (see Appendix). Introducing new variables $x=m/T$ and $y=p/T$
and the new unknown function $C(x,y)= \exp(\sqrt{x^2+y^2})f_m(x,y)$
and assuming that the temperature drops with the expansion in accordance with
the usual law, $\dot T=-HT$, we can rewrite this equation in the form:
\begin{eqnarray}
Hx\partial_x (Ce^{-u}) = {(-|A_0|^2)me^{-u/2} \over 64\pi^3 uxy }
\int_0^\infty {dy' y' e^{-u/2} \over u' } \nonumber \\
\left[ C(x,y) - C(x,y')\right] \left[ e^{-|y-y'|/2} - e^{(y+y')/2}\right]
\label{hxdf}
\end{eqnarray}
where $u =\sqrt{x^2+y^2}$ and $u' =\sqrt{x^2+y'^2}$.

Successively multiplying eq.(\ref{hxdf}) by $\exp[(u+y)/2]$ and by $\exp(-y)$
and each time differentiating in $y$  we arrive after some
straightforward algebra to the following {\it differential} equation:
\beq{
 JC'' + 2 J'C' =
- { 64\pi^3 H x^2 \over |A_0|^2 m} e^{y/2} \partial_y \left\{ e^{-y}
\partial_y\left[ e^{(u+y)/2} uy\partial_x \left(Ce^{-u}\right)\right]\right\}
\label{jc''}
}\eeq
where prime means differentiation with respect to $y$ and $J$ is given by the
expression:
\beq{
 J(x,y)= {1\over 2} e^{y/2} \int^\infty_{u+y} dz e^{-z/2}
\left( 1- {x^2\over z^2}\right)-
{1\over 2} e^{-y/2} \int^\infty_{u-y} dz e^{-z/2}
\left( 1- {x^2\over z^2}\right)
\label{jxy }
}\eeq
It can be checked that $J$ satisfies the differential equation:
\beq{
 J'' - J/4 = - ye^{-u/2}/u
\label{j''}
}\eeq

We will solve eq.(\ref{jc''}) perturbatively, looking for the solution in the
form $C(x,y) = C_0(x) + C_1(x,y)$ where $C_0(x)$ is the equilibrium solution;
it can be found from the condition of particle number conservation:
\beq{
C_0(x) \int_0^\infty dy y^2 e^{-u(x,y)} \sim 1/x^3
\label{c0}
}\eeq
If the rate of the elastic scattering is large in comparison with the rate of
the universe expansion, or in other words, the dimensionless coefficient
in front of the r.h.s. of eq.(\ref{jc''}) is sufficiently small we can neglect
$C_1$ in the r.h.s. and explicitly solve
this equation in quadratures.  Twice integrating by parts we get:
\begin{eqnarray}
C'_1 (x,y) = {64\pi^3 C_0(x)\over |A_0|^2 } {Hx^2 \over m }
\left\{ {e^{-u(x,y)/2} \over J} \left[ -{\partial_x C_0 \over C_0}
\left( u(x,y)
+{y^2\over u(x,y)} - {y^2\over 2} \right) \right. \right. \nonumber \\
- \left. {xy^2 \over 2u(x,y)} + x
+ u(x,y)y{J'\over J}\left( {\partial_x C_0 \over C_0} -
{x\over u(x,y)} \right) \right] \nonumber \\
\left. -{1\over J^2}
\int_y^\infty dzz^2 e^{-u(x,z)}\left[ {\partial_x C_0 \over C_0}
-{x\over u(x,z)} \right] \right \}
\label{c'}
\end{eqnarray}

The function $C_1(x,y)$ is found by the integration of this expression in $y$
and the integration constant (which is a function of $x$) should be chosen from
the condition of the particle number conservation:
\beq{
\int dy y^2 e^{-u(x,y)} C_1(x,y) =0
\label{intc1}
}\eeq

In a simple case of the universe dominated by relativistic particles the
Hubble parameter is given by the expression
\beq{
 H= \sqrt{{4\pi^3 g_*\over 45}} {m^2 \over m_{Pl}x^2}
\label{hub}
}\eeq
where $m_{Pl}$ is the Planck mass and $g_*$ is the number of relativistic
degrees of freedom contributing into the cosmological energy density. Thus
$Hx^2$ is $x$-independent and it is convenient to consider the relative
quantity $r'_0 = (C_1'/C_0) (m/Hx^2) (|A_0|^2 / 64\pi^3)$. It is presented in
fig.1 for  $x=10$ as a function of $y$ and
in fig.2 for $y=3$ as a function of $x$.
With rising $x$ the correction
$C_1(x,y)$ is getting bigger; asymptotically for large $x$, $C_1 \sim x$. We
can trust these results till $C_1/C_0 \ll 1$ at least in the region of $y$
where the supression due to the factor $\exp[-u(x,y)]$ is not too strong, i.e.
for $y^2 \sim 2x$.

\section{Power series expansion for the case of a constant amplitude.}

Here we will present an alternative approach to the solution of the original
kinetic equation (\ref{hxdf}). This equation is not as easy to solve
perturbatively (for small deviations from kinetic equilibrium) as the
equivalent differential equation (\ref{dtfm}). However the method considered
here can be useful for the solution of the kinetic equation for the case
of weak interactions when the amplitude strongly depends on the energies of
the colliding particles and the reduction to the more simple form (\ref{dtfm})
is not found. Below we will find the solution of eq.(\ref{hxdf}) expanding
$C(x,y)$ in powers of $y$ (in fact of $y^2$). Formally the expansion is valid
even if the elastic scattering is relatively weak and the deviations from the
kinetic equilibrium are large, but the convergence of the expansion is getting
worse for larger $x$, when the probability of elastic scattering goes down.

After a simple rearrangement we can rewrite eq.(\ref{hxdf}) in the form:
\begin{eqnarray}
{64\pi^3 Hx^2 C_0(x) y\exp(-u/2)\over |A_0|^2 m }
\left(x-{u\partial_x C_0(x) \over C_0(x)} \right)= \nonumber \\
C_1(x,y)J(x,y) -
2\sinh (y/2) \int^\infty_0{dy'y'\over u'} e^{-(u'+y')/2} C_1(x,y') \nonumber \\
+2\int^y_0{dy'y'\over u'} e^{-u'/2} C_1(x,y')\sinh [(y-y')/2]
\label{pi3}
\end{eqnarray}
We will look for the solution of this equation in the form of the Taylor
expansion
\beq{
 C_1(x,y) =C_{10}(x) +  C_{12}(x) y^2/2 + C_{14}(x) y^4/4 + ...
\label{c1xy}
}\eeq
The coefficient $C_{10}$ is not determined by the equation but should be found
from the condition of conservation of the particle number as is discussed
above.
One can easily check that the odd powers of $y$ in this expansion are absent.
This was evident if fact from the very beginning because the expansion in terms
of the three momentum $| \vec p |$ normally goes in terms of $\vec p\,^2$ and
there is no reason to expect a singularity $\sqrt{ \vec p\,^2}$
in the considered problem.

We will retain in $C$ only terms up to the fourth power in $y$, so the
last integral in equation  (\ref{c1xy}) which contributes in the order $y^5$
can be neglected.  After a simple algebra the following system of equations,
determining $C_{12}$ and $C_{14}$, can be obtained:
\beq{
{1\over 2} C_2 I_3 + {1\over 4} C_4 I_5 = x\left( {\partial_x C_0 \over C_0}
                                         - 1\right)
\label{c2i3}
}\eeq
\beq{
{1\over 2} C_2 I_1 - \left({1\over 4x} +{1\over 24} \right) \left(
{1\over 2} C_2 I_3 + {1\over 4} C_4 I_5 \right)
= - {\partial_x C_0 \over 2xC_0}
\label{sys0}
}\eeq
where
\beq{
I_n(x) = \int^\infty_0 {dy' y'^n \over u'} e^{-(u'-x+y')/2}
\label{in}
}\eeq

For example for $x=10$ we find $C_2 = -1$ and $C_4 = 0.007$. It very well
agrees with the results of the previous section up to $y \approx 10$. At these
and larger $y$'s the suppression of the spectrum due to the factor $\exp(-u)$
is already very strong and this part of the spectrum is not physically
significant. It is noteworthy that there is no formal small parameter in the
equation (\ref{c2i3}) and all $C_n$ would in principle equally
contribute. However higher order terms are small numerically, possibly
because of nonrelativistic nature of the problem, and thus  the neglect of
those
terms is justified. For example if we neglect the term $C_4 y^4$
(which we did not) the resulting value of $C_2$ would be only slightly
different form the one found together with $C_4$.

\section{Weak interaction scattering.}

Now let us consider a more realistic case of elastic scattering of  Dirac
massive tau-neutrino on massless $\nu_e$, $\nu_\mu$, and $e^\pm$, which
are in thermal equilibrium. The amplitude of elastic
scattering summed over all light lepton species is
\beq{
 |A_w|^2 = K [(pk)^2 + (pk')^2]/m^4
\label{aw}
}\eeq
where (for $\sin \theta_w =0.23$)
\beq{
K=32G_F^2 m^4 (3-4\sin^2 \theta_w + 8\sin^4 \theta_w) \approx
1.09\times 10^{-20} (m/ MeV)^4
\label{k}
}\eeq

We have to substitute the amplitude $A_w$ instead of $A_0$ into
eq.(\ref{dtfm}).
The final result of the integration is grossly simplified in nonrelativistic
limit when one neglects the terms of the order of $p^2/m^2$ in the amplitude.
The details of the integration can be found in the Appendix. Performing it we
find, instead of eq.(\ref{hxdf}), the following equation:
\begin{eqnarray}
 Hx\partial_x (Ce^{-u}) =
-{12Kmue^{-u/2} \over 64\pi^3 y x^5}
\int_0^\infty {dy' y' \over u'}e^{-(u'+y')/2}[C(x,y)-C(x,y')]\nonumber \\
\left[ 2\sinh(y/2) (1+y'/3 + (y^2+y'^2)/24) - 2y \cosh(y/2) (1/3+y'/12)\right]
 \label{hxdfw}
\end{eqnarray}
As in the previous sections we make the perturbative expansion $C(x,y)=C_0(x)
+C_1(x,y)$ and expand $C_1$ in powers of $y$. In eq.(\ref{hxdfw})
we have  kept only the terms which
are essential for calculations of $C_1$ up to $y^4$. The exact r.h.s. of this
equation is presented in the Appendix.

We will slightly change notations here and expand the relative deviation
from the equilibrium distribution:
 \beq{
 r_w(x,y)=C_1(x,y)/C_0(x) = r_2(x) y^2 /2 + r_4(x) y^4/4
\label{rxy}
}\eeq
For the coefficients $r_n$ the following equations are valid:
\beq{
{1\over 2}r_2 F_3  + {1\over 4}r_4 F_5 =
S\left( {\partial_x C_0 \over C_0}-1\right)
\label{r2f3}
}\eeq
\beq{
{r_2 \over 2}\left[ F_1 +L_3 - \left( {1\over 4x} + {1\over 2x^2}\right)
F_3 \right]
+{r_4\over 4}\left[ L_5 - \left( {1\over 4x} + {1\over 2x^2}\right)F_5
\right] = {-S\over 2x^2}
\label{r2f1}
}\eeq
Here $S=16\pi^3 (Hx^2)x^4 /Km^4$ and
\beq{
 F_n=\int_0^\infty dy y^n {x\over u} \left(1+{y\over 2} + {y^2 \over 8}\right)
e^{(u-x+y)/2}
\label{fn}
}\eeq
\beq{
L_n=\int_0^\infty dy y^n {x\over u} \left({y\over 48} - {y^2 \over 192}\right)
e^{(u-x+y)/2}
\label{ln}
}\eeq

For example for $m=20$ MeV and $x=10$ we get $r_2=-4.5\times 10^{-2}$ and
$r_4 = 6.66\times 10^{-4}$. Behavior of $r_w(x,y)$ for $x=10$ in units
$S/x^4 \approx 18.6 {\rm{Mev}}^3/m^3 $ is presented in fig.3. The coefficients
$r_2(x)$ and $r_4(x)$ in the same units are presented in figs.4 and fig.5.
The corrections quickly rise with $x$ (remember the factor $x^4/S$) but they
remain reasonably small till the freezing of the annihilation of tau neutrinos.

Comparing the annihilation rate for $\nu_\tau$ in
kinetic equilibrium with the rate
of annihilation of $\nu_\tau$ with the spectrum calculated here, we conclude
that the frozen number density of nu-tau can be bigger by about 10\% than that
found in the standard case.  However, calculating the spectrum,
we have neglected the annihilation of $\bar\nu_\tau \nu_\tau$ and this
process may have an essential influence on the spectrum and on the final result
on the frozen abundance. We will take this into account in the subsequent work.

\bigskip
{\bf Acknowledgment.}
This paper was supported in part by the Danish National Science Research
Council
through grant 11-9640-1 and in part by Danmarks Grundforksningsfond through its
support of the Theoretical Astrophysical Center.

\appendix
\section{Appendix.}

Integration in the r.h.s. of eq.(\ref{dtfm}) can be done in the following
direct
way. First using $\delta^3$-function we integrate over $d^3k$ so that
$\vec k = \vec k' - \vec p + \vec p'$. Correspondingly $\omega^2 = \vec k^2 =
\omega'^2 - 2\omega' R \cos \theta_k + R^2$ where $ R = |\vec p -\vec p'|$ and
$\theta$ is the angle between $\vec R$ and $\vec k'$.

Since $d\cos\theta$, which comes from $d^3k'$, can be whiten as
$$
 d\cos \theta_k = - d\omega\omega /\omega' R
\eqno{(A1)}
$$
we can easily integrate over $d\omega$ using
$\delta(\omega -\omega ' + E -E') $. The limits of the integration
over $\omega'$ are determined by the condition $|\cos \theta | <1$ so we get
$$
 \int {d^3k\over \omega} {d^3 k' \over \omega'} \delta^4(p+k-p'-k') =
{2\pi \over R} \int^\infty_{\omega_{min}} d\omega' \langle ... \rangle_{\phi_k}
\eqno{(A2)}
$$
where $\omega_{min}= (R+E-E')/2$ and the expression in the angle brackets
means averaging over the polar angle of the vector $\vec k'$. For a constant
amplitude there is no dependence on this angle.

In the remaining integration over $d^3p'$ we change the integration over the
azimuthal angle to the integration over $R$ in accordance with the relation
$d\cos\theta_p = - RdR / p p'$. Correspondingly we get
$$
\int {d^3 p' \over E'} = (2\pi) \int^\infty_0 {dp'p' \over pE'}
\int^{R_{max}}_{R_{min}} dR \langle ...\rangle_{\phi_p}
\eqno{(A3)}
$$
where, as above the averaging over $p'$-polar angle is done. The limits of
integration in $R$ are $R_{min}= |p-p'|$ and $R_{max}=p+p'$.

If the amplitude of the scattering is constant and if the light particle
distribution functions are known, $f_0 =\exp(-E/T)$, we can easily make
all the integration but one (in $dp'$) and obtain eq.(\ref{hxdf}).
For the case of weak interactions the amplitude (\ref{aw}) is a function of
momenta and the integration is more complicated. We will make it in the
nonrelativistic limit, neglecting terms of the order of $v^2 = p^2/E^2$ in the
amplitude. In this limit $|A_w|^2 = K(E/m)^2((\omega^2+\omega'^2)/m^2$. The
absence of the angular dependence in the amplitude simplifies the calculations
very much. Performing the integration in the way described above, we obtain for
the r.h.s. of eq.(\ref{dtfm}) (with the substitution $A_0\rightarrow A_w$)
the following expression:
\beq{
(r.h.s.)={KE \over 128\pi^3 p m^4}\int_0^\infty{dp'p'\over E'}
\int^{p+p'}_{|p-p'|}dR\int^\infty_{\omega_{min}}d\omega' e^{(\omega'+E')/T}
(\omega^2+\omega'^2) [C(p')-C(p)]
\label{rhs}
}\eeq
where $\omega =\omega' +E-E'$.
Making a trivial integration over $\omega$ and neglecting terms proportional to
$(E-E')^2 \sim m^{-4}$ and also neglecting the integral similar to the last one
in eq.(\ref{pi3}) we arrive to eq.(\ref{hxdfw}).

\newpage

{\bf Figure captions.}

{\bf Fig.1.} Relative deviation from the thermal equilibrium of the
normalized $y$-derivative of $C_1(x,y)$ for the case of
a constant amplitude of elastic scattering
for fixed $x=10$ and running $y$, $r'(10,y)$. See the definition of
$r'$ at the end of Sec. 2.

{\bf Fig.2.} The same as in Fig.1 but for fixed $y=3$ and
running $x$, $r'(x,3)$.

{\bf Fig.3.} The correction to the spetral density of nu-tau for $x=10$,
$r_w(10,y)$ in units $S/x^4$ (see the end of Sec.4).

{\bf Fig.4.} Taylor expansion coefficient $r_2(x)$ in units $S/x^2$.

{\bf Fig.5.} Taylor expansion coefficient $r_4(x)$ in units $S/x^2$.

\end{document}